\documentclass[prc,superscriptaddress,twocolumn,preprintnumbers,showpacs,nofootinbib,review]{revtex4}

\usepackage{graphicx}% Include figure files
\usepackage{dcolumn}% Align table columns on decimal point
\usepackage{bm}% bold math
\usepackage{color}%
\graphicspath{{../Figures-EPS/}{}{.}}
\usepackage{verbatim}

\begin{document}

\title{The $^{14}$N(p,$\gamma$)$^{15}$O reaction studied with a composite germanium detector}

\author{M.~Marta}\thanks{Present address: GSI Helmholtzzentrum f\"ur Schwerionenforschung, Darmstadt, Germany}\affiliation{Helmholtz-Zentrum Dresden-Rossendorf, Bautzner Landstr. 400, 01328 Dresden, Germany}\thanks{As of 1 January 2011, Forschungszentrum Dresden-Rossendorf (FZD) has been renamed to Helmholtz-Zentrum Dresden-Rossendorf (HZDR).}
\author{A.~Formicola}\affiliation{INFN, Laboratori Nazionali del Gran Sasso (LNGS), Assergi (AQ), Italy}
\author{D.~Bemmerer}\affiliation{Helmholtz-Zentrum Dresden-Rossendorf, Bautzner Landstr. 400, 01328 Dresden, Germany}
\author{C.~Broggini}
 \affiliation{Istituto Nazionale di Fisica Nucleare (INFN), Sezione di Padova, via Marzolo 8, 35131 Padova, Italy}
\author{A.~Caciolli}\affiliation{Istituto Nazionale di Fisica Nucleare (INFN), Sezione di Padova, via Marzolo 8, 35131 Padova, Italy}\affiliation{Dipartimento di Fisica, Universit\`a degli studi di Siena, Italy}
\author{P.~Corvisiero}\affiliation{Universit\`a di Genova and INFN Sezione di Genova, Genova, Italy}
\author{H.~Costantini}\affiliation{Universit\`a di Genova and INFN Sezione di Genova, Genova, Italy}
\author{Z.~Elekes}\affiliation{Institute of Nuclear Research (ATOMKI), Debrecen, Hungary}
\author{Zs.~F\"ul\"op}\affiliation{Institute of Nuclear Research (ATOMKI), Debrecen, Hungary}
\author{G.~Gervino}\affiliation{Dipartimento di Fisica Sperimentale, Universit\`a di Torino and INFN Sezione di Torino, Torino, Italy}
\author{A.~Guglielmetti}\affiliation{Universit\`a degli Studi di Milano and INFN, Sezione di Milano, Italy}
\author{C.~Gustavino}\affiliation{INFN, Laboratori Nazionali del Gran Sasso (LNGS), Assergi (AQ), Italy}
\author{Gy.~Gy\"urky}\affiliation{Institute of Nuclear Research (ATOMKI), Debrecen, Hungary}
\author{G.~Imbriani}\affiliation{Dipartimento di Scienze Fisiche, Universit\`a di Napoli ''Federico II'', and INFN Sezione di Napoli, Napoli, Italy}
\author{M.~Junker}\affiliation{INFN, Laboratori Nazionali del Gran Sasso (LNGS), Assergi (AQ), Italy}
\author{A.~Lemut}\affiliation{Universit\`a di Genova and INFN Sezione di Genova, Genova, Italy}
\author{B.~Limata}\affiliation{Dipartimento di Scienze Fisiche, Universit\`a di Napoli ''Federico II'', and INFN Sezione di Napoli, Napoli, Italy}
\author{C.~Mazzocchi}\affiliation{Universit\`a degli Studi di Milano and INFN, Sezione di Milano, Italy}
\author{R.~Menegazzo}\affiliation{Istituto Nazionale di Fisica Nucleare (INFN), Sezione di Padova, via Marzolo 8, 35131 Padova, Italy}
\author{P.~Prati}\affiliation{Universit\`a di Genova and INFN Sezione di Genova, Genova, Italy}
\author{V.~Roca}\affiliation{Dipartimento di Scienze Fisiche, Universit\`a di Napoli ''Federico II'', and INFN Sezione di Napoli, Napoli, Italy}
\author{C.~Rolfs}\affiliation{Institut f\"ur Experimentalphysik III, Ruhr-Universit\"at Bochum, Bochum, Germany}
\author{C.~Rossi~Alvarez}\affiliation{Istituto Nazionale di Fisica Nucleare (INFN), Sezione di Padova, via Marzolo 8, 35131 Padova, Italy}
\author{E.~Somorjai}\affiliation{Institute of Nuclear Research (ATOMKI), Debrecen, Hungary}
\author{O.~Straniero}\affiliation{Osservatorio Astronomico di Collurania, Teramo, and INFN Sezione di Napoli, Napoli, Italy}
\author{F.~Strieder}\affiliation{Institut f\"ur Experimentalphysik III, Ruhr-Universit\"at Bochum, Bochum, Germany}
\author{F.~Terrasi}\affiliation{Seconda Universit\`a di Napoli, Caserta, and INFN Sezione di Napoli, Napoli, Italy}
\author{H.P.~Trautvetter}\affiliation{Institut f\"ur Experimentalphysik III, Ruhr-Universit\"at Bochum, Bochum, Germany}
\author{A.~Vomiero}\affiliation{Consiglio nazionale delle ricerche, CNR-IDASC Sensor Lab, Brescia, Italy}

\collaboration{The LUNA Collaboration}\noaffiliation

\begin{abstract}
The rate of the carbon-nitrogen-oxygen (CNO) cycle of hydrogen burning is controlled by the $^{14}$N(p,$\gamma$)$^{15}$O reaction. The reaction proceeds by capture to the ground states and several excited states in $^{15}$O. In order to obtain a reliable extrapolation of the excitation curve to astrophysical energy, fits in the R-matrix framework are needed. 
In an energy range that sensitively tests such fits, new cross section data are reported here for the four major transitions in the $^{14}$N(p,$\gamma$)$^{15}$O reaction. 
The experiment has been performed at the Laboratory for Underground Nuclear Astrophysics (LUNA) 400\,kV accelerator placed deep underground in the Gran Sasso facility in Italy. Using a composite germanium detector, summing corrections have been considerably reduced with respect to previous studies. 
The cross sections for capture to the ground state and to the 5181, 6172, and 6792\,keV excited states in $^{15}$O have been determined at 359, 380, and 399\,keV beam energy. In addition, the branching ratios for the decay of the 278\,keV resonance have been remeasured. 

\end{abstract}

\pacs{25.40.Ep, 25.40.Lw, 26.20.Cd, 26.65.+t}

\maketitle

\section{Introduction}

The stellar rate of the carbon-nitrogen-oxygen (CNO) cycle of hydrogen burning \cite{Bethe39-PR_letter,Weizsaecker38-PZ} is controlled by the slowest process, the $^{14}$N(p,$\gamma$)$^{15}$O reaction \cite{Iliadis07-Book}. In the Sun, hydrogen burning proceeds mainly by the competing proton-proton chain, and the CNO cycle contributes only 0.8\% to the energy production \cite{Bahcall05-ApJL}. However, solar CNO hydrogen burning gives rise to neutrino emission lines from the $\beta^+$ decay of $^{13}$N and $^{15}$O \cite{Bahcall05-ApJL}. It has recently been suggested \cite{Haxton08-ApJ} to use the expected CNO neutrino flux data from the Borexino detector \cite{Borexino08-PRL} and the planned SNO+ \cite{Chen05-NPBPS} detector to measure the abundance of carbon and nitrogen in the solar core. This would address the so-called solar metallicity problem \cite{PenaGaray08-arxiv,Serenelli09-ApJL}, which is given by the fact that the new solar metallicities \cite{Asplund09-ARAA} lead to inconsistencies in the standard solar model. The correct interpretation of the expected CNO neutrino data requires, however, that the nuclear reaction rate of the CNO cycle, which is determined by the $^{14}$N(p,$\gamma$)$^{15}$O cross section, be known with sufficient precision.

The $^{14}$N(p,$\gamma$)$^{15}$O cross section $\sigma(E)$ can be parameterized using the astrophysical S-factor
\begin{equation}
S(E) = \sigma E \exp \left[212.4/\sqrt{E} \right],
\end{equation}
with $E$ denoting the energy in the center of mass system in keV. 

\begin{figure}[tb]
\centering
 \includegraphics[angle=00,width=1.0\columnwidth]{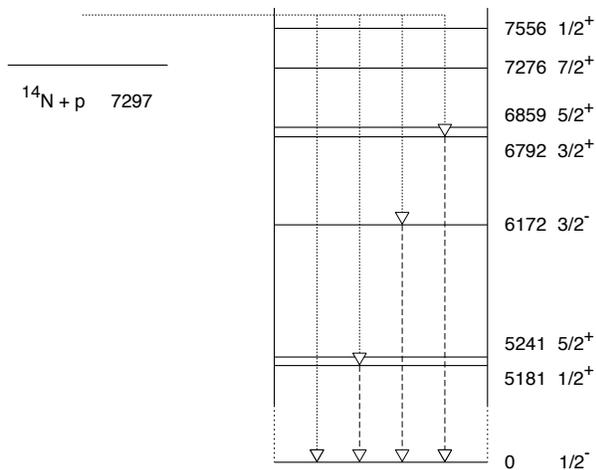}
 \caption{\label{fig:Levels-O15} Level scheme of $^{15}$O, in keV \cite{Ajzenberg91-NPA,Imbriani05-EPJA}. The most important $\gamma$-transitions are denoted by arrows.}
\end{figure}

The excitation function has been studied previously \cite[e.g.]{Lamb57-PR,Hebbard63-NP,Schroeder87-NPA}, and these data determine the recommended value in the current nuclear reaction rate compilations for astrophysics \cite{CF88-ADNDT,Adelberger98-RMP,NACRE99-NPA}. Subsequently, a number of new experimental and theoretical results on this reaction have been reported \cite{Bertone01-PRL,Angulo01-NPA,Mukhamedzhanov03-PRC,Yamada04-PLB,Formicola04-PLB,Runkle05-PRL,Imbriani05-EPJA,Schuermann08-PRC,Marta08-PRC,Marta10-PRC}, showing that the recommended value of the reaction rate \cite{CF88-ADNDT,Adelberger98-RMP,NACRE99-NPA} has to be revised downward by a factor of two. In particular, capture to the ground state in $^{15}$O (fig.~\ref{fig:Levels-O15}) was shown to be strongly suppressed \cite{Bertone01-PRL,Angulo01-NPA,Mukhamedzhanov03-PRC,Yamada04-PLB,Formicola04-PLB,Runkle05-PRL,Imbriani05-EPJA}. This reduction is now adopted in a very recent compilation \cite{Seattle09-workshop}. However, some open questions remain. 

In particular, two groups have in recent years presented cross section data and performed R-matrix fits based on their new data: LUNA \cite{Formicola04-PLB,Imbriani05-EPJA} and TUNL \cite{Runkle05-PRL}. These two works show excellent agreement when it comes to the most important contribution to the total S-factor, namely capture to the state at 6792\,keV: 1.20$\pm$0.05~keV\,barn \cite{Imbriani05-EPJA} and 1.15$\pm$0.05~keV\,barn \cite{Runkle05-PRL}, respectively. However, their results differ by much more than the quoted uncertainties when it comes to the second most important contribution, capture to the ground state: Whereas LUNA reported%
\footnote{$S_i(0)$ denotes the S-factor, extrapolated to zero energy, for capture to the state at $i$~keV in $^{15}$O. $S_{\rm GS}(0)$ and $S_{\rm tot}(0)$ refer to ground state capture and to the total S-factor, respectively.}
$S_{\rm GS}$(0) = 0.25$\pm$0.06\,keV\,barn \cite{Formicola04-PLB}, the TUNL value is double that, 0.49$\pm$0.08\,keV\,barn \cite{Runkle05-PRL}. This discrepancy amounts to about 15\% of the total extrapolated $S_{\rm tot}(0)$, dominating the uncertainty. 

The only significant methodical difference between the two fits from LUNA \cite{Formicola04-PLB,Imbriani05-EPJA} and TUNL \cite{Runkle05-PRL}  is the treatment of high-energy data. The LUNA fit is a global fit, based on the LUNA data presented in the same paper, and on the Schr\"oder {\it et al.} data \cite{Schroeder87-NPA} which had been corrected for the summing-in effect. The TUNL fit, on the other hand, is a partial fit based solely on the TUNL data presented in the same paper, with the higher-energy R-matrix poles kept fixed based on a previous fit of the Schr\"oder {\it et al.} data \cite{Schroeder87-NPA}. The starting values and general procedure for both fits are otherwise the same \cite{Angulo01-NPA}. 

The experimental data points by LUNA \cite{Formicola04-PLB,Imbriani05-EPJA} and TUNL \cite{Runkle05-PRL} are generally in agreement with each other, but they show some systematic uncertainty due to the fact that both groups had employed large germanium detectors in close geometry. This arrangement had been chosen in order to obtain a high enough detection efficiency for the weak ground state capture line. However, in this way both groups also incurred true coincidence summing-in corrections of more than 100\% for the ground state data. Such a large correction entails considerable systematic uncertainty.

% ==================
% ==================

The aim of the present work is to address the conflicting extrapolations \cite{Formicola04-PLB,Runkle05-PRL} in two ways. The experimental problem of the previous high summing-in correction is solved by using a Clover detector. The problem of the selection of the database is solved by providing the ground state cross section relative to that for the well-known capture to the state at 6792\,keV. The present relative data can then be added to one particular data set without introducing additional scaling uncertainty. Alternatively, they can be rescaled to absolute data using an overall fit of 6792\,keV capture based on several independent works, strongly reducing the scaling uncertainty. 

For the present experiment, the energy range of $E$ = 317-353\,keV has been selected, far enough above the 259\,keV resonance to limit resonant contributions, and at the same time a region where a sensitive minimum \cite{Angulo01-NPA}
of R-matrix fits is observed. In principle, such a measurement would also have been possible at $E$ $\approx$ 170\,keV, in a second sensitive minimum. However, the yield is a factor 100 lower there, so that the present energy range was chosen for practical purposes.

The present relative cross section data have been published previously in abbreviated form \cite{Marta08-PRC}. The present work provides full details of that experiment and analysis. In addition, new branching ratios for the decay of the 259\,keV resonance obtained in even farther geometry are presented here. The absolute off-resonance $^{14}$N(p,$\gamma$)$^{15}$O cross section for capture to the ground state and the 5181, 6172, and 6792\,keV excited states is derived at $E_{\rm p}$ = 359, 380, and 399\,keV. In order to improve the reliability, this latter analysis is performed in two independent ways, namely by the $\gamma$-line shape method \cite{Formicola03-NIMA} and by the classical peak integral approach.

% ===========================================================
\section{Experiment}

% ================
\begin{figure}[tb]
\centering
 \includegraphics[angle=90,width=0.7\columnwidth]{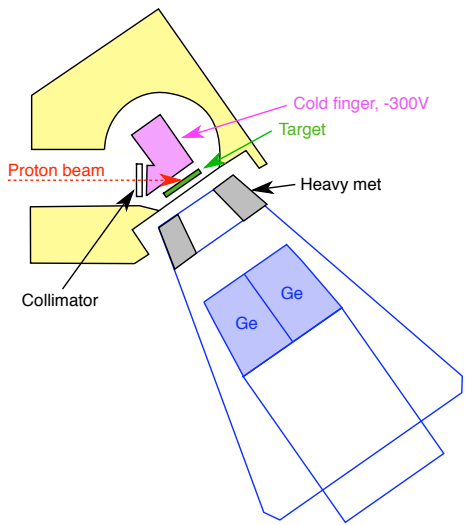}
 \caption{\label{fig:Cloversetup}(Color online) Schematic view of the experimental setup.}
\end{figure}

The experiment was performed in the Laboratory for Underground Nuclear Astrophysics (LUNA) at the Gran Sasso National Laboratory (Italy) \cite{Costantini09-RPP,Broggini10-ARNPS}. 
At the LUNA site, the $\gamma$-ray laboratory background for $E_\gamma$ $>$ 3\,MeV is strongly reduced due to the rock overburden equivalent to 3800 meters water \cite{Bemmerer05-EPJA,Szucs10-EPJA}. Also for $E_\gamma$ $\leq$ 3\,MeV with proper shielding the $\gamma$-ray background has been found to be much lower than in comparable laboratories at the surface of the Earth \cite{Caciolli09-EPJA}. The unique location of LUNA has enabled the study of several nuclear reactions of astrophysical importance \cite[]{Bonetti99-PRL,Formicola04-PLB,Imbriani04-AA,Lemut06-PLB,Bemmerer06-PRL,Bemmerer09-JPG,Limata10-PRC}.

\subsection{Setup}
\label{subsec:Setup}

The LUNA2 400\,kV accelerator \cite{Formicola03-NIMA} provided a H$^+$ beam of $E_{\rm p}$ = 359, 380, and 399\,keV, with 0.25-0.45\,mA intensity. The ion beam passed a collimator of 5\,mm diameter, which absorbed a few percent of the full beam intensity, and a cold trap cooled by liquid nitrogen (fig.~\ref{fig:Cloversetup}), before hitting the target. Secondary electrons emitted from the target surface were suppressed by applying -~300\,V suppression voltage to the cold trap. 
The reproducibility of the current from run to run is estimated to be 2\%.

% ================
\begin{figure}[]
\centering
\includegraphics[angle=0,width=1.0\columnwidth]{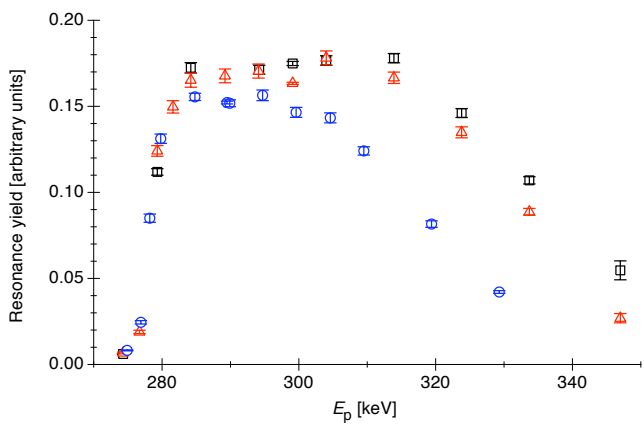}
 \caption{\label{fig:Clover_targetscans}(Color online) Scans of the $E_{\rm p}$ = 278\,keV resonance at the beginning of the experiment (black squares), one day with 29\,C dose later (red triangles), and at the end of the experiment after a total accumulated dose of 267\,C (blue circles).}
\end{figure}
% ================

\subsection{Target}
\label{subsec:Target}

A titanium nitride target produced by reactive sputtering at Laboratori Nazionali di Legnaro was used for the experiments. The target had 60\,keV energetic width at the $E_{\rm p}$ = 278\,keV resonance (fig.~\ref{fig:Clover_targetscans}), when irradiated under 55$^\circ$ angle. In order to obtain its stoichiometry, the stopping power at the resonance energy, the beam current and the strength of the monitor resonance must be known. For the stopping of protons in titanium and nitrogen, the values from the ${\rm SRIM}$ software \cite{SRIM08.02} have been used. For the strength of the resonance, $\omega\gamma$ = 13.1$\pm$0.6\,meV was adopted, the recommended value from Ref.~\cite{Seattle09-workshop}.
Based on this number, a stoichiometric ratio Ti:N of 1:0.93 has been determined. 
The target stoichiometry gives rise to 6\% systematic uncertainty in the absolute cross section results, mainly from the reference $\omega\gamma$ value. 

In order to properly correct for the change of the target under intense proton bombardment, during the experiment the target profile was monitored every day by scanning the $E_{\rm p}$ = 278\,keV resonance (fig.~\ref{fig:Clover_targetscans}). The sharp low-energy edge of the profile is given by the convolution of the 0.1\,keV energy spread of the beam \cite{Formicola03-NIMA} and the 1.06\,keV natural width of the resonance \cite{Ajzenberg13-15}. On the ensuing constant plateau, the step height is proportional to the inverse of the effective stopping power per $^{14}$N nucleus in the compound.

A reduction of up to 7\% in the integral of the target profile was observed from day to day, with a typical proton dose of 24\,C (1.5 $\cdot$ 10$^{20}$ H$^+$ ions) deposited on the target per day. 
It is estimated that the target composition is known with 5\% precision for any given time during the experiment.

\subsection{Detection of emitted $\gamma$-rays}
\label{subsec:GammaDetection}

% ================
\begin{figure}[tb]
\centering
 \includegraphics[angle=-90,width=1.0\columnwidth]{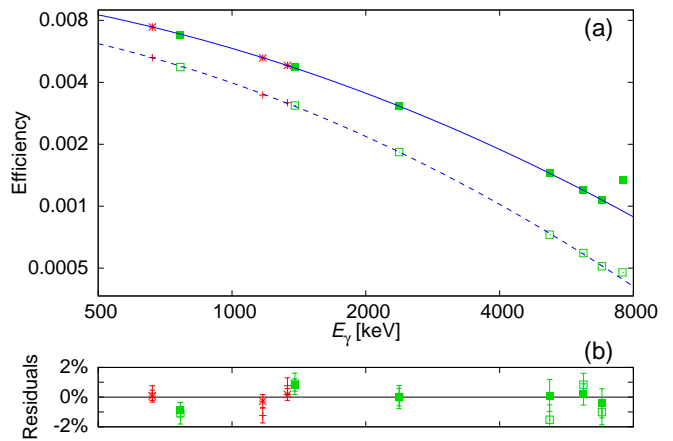}
 \caption{\label{fig:Efficiency}(Color online) (a) $\gamma$-ray detection efficiency for the detector at 9.5\,cm distance from the target, as determined with radioactive sources and the two-line method at the 259\,keV resonance. Solid (dashed) curve, efficiency for addback (singles) mode. (b) Residuals. The data point at 7556\,keV is not yet corrected for summing-in. It was excluded from the fit, and is shown here for illustration only. The pair of $\gamma$-rays at 5181 and 2375\,keV was also not included in the fit but is plotted here as a check on the reliability of the curve.}
\end{figure}
% ================

The $\gamma$-rays emitted from the target were detected in a Eurisys Clover-BGO detection system \cite{Elekes03-NIMA} placed at an angle of 55$^\circ$ with respect to the beam axis. The front end of the Clover detectors was at 9.5\,cm distance from the target. For the branching-ratio measurement (sec.~\ref{subsec:Branchings}), the front end was placed at 19.5\,cm distance from the target instead. 

The output signal from each of the four Clover crystals was split into two branches called branch 'S' and branch 'A'. For branch 'S', each of the four signals was amplified and digitized separately, and the four spectra were gainmatched and summed in the offline analysis, giving the so-called singles mode. 

For branch 'A', the preamplifier output signals were gain-matched and added in a homemade analog summing unit. The added signal was then amplified and digitized, giving the so-called addback mode spectra.  Typical resolutions for addback (singles) mode were 9\,keV (3.3\,keV) at 1.3\,MeV and 12\,keV (6\,keV) at 6.8\,MeV. For experiments off the 259\,keV resonance, the addback mode data were recorded in anticoincidence with the BGO escape-suppression shield to reduce the Compton background.

The $\gamma$-ray detection efficiency was measured using $^{137}$Cs and $^{60}$Co radioactive sources calibrated to 1.5\% and 0.75\% (1$\sigma$ confidence range), respectively. The efficiency curve (fig.~\ref{fig:Efficiency}, upper panel) was then extended to high energy based on spectra recorded at the 259\,keV 1/2$^+$ resonance, using the known 1:1 $\gamma$-ray cascades for the excited states at 6172 and 6792\,keV \cite{Ajzenberg13-15}. The $\gamma$-rays from the decay of this 1/2$^+$ resonance are isotropic \cite{Ajzenberg13-15}. The angular correlations of 8-10\% between primary and secondary $\gamma$-ray are experimentally well known \cite{Povh59-PR}. They result in up to 0.4\% correction on the efficiency curve, because they affect the summing-out correction. For the worst case, the 6792\,keV $\gamma$-ray, the calculated summing-out correction is 3.6\% in addback mode (1.1\% in singles mode), with an assumed relative uncertainty of 20\%. This result is consistent with a GEANT4 \cite{Agostinelli03-NIMA} simulation showing (4.5$\pm$1.8)\% correction. 

As a check on the quality of the efficiency curve, the experimental cascade ratio for the 5181\,keV excited state (not used in the fit) was found to be reproduced within 1\% statistics (fig.~\ref{fig:Efficiency}, lower panel), again assuming 1:1 $\gamma$-ray cascade ratio \cite{Ajzenberg13-15}.

% ===========================================================
\section{Data analysis and results}

The first part of the analysis concentrated on the ratio of the cross sections for radiative proton capture to the ground state and the fourth excited state at 6792\,keV in $^{15}$O, determined with the detector at 9.5\,cm distance from the target. 
These relative data have been reported previously in abbreviated form \cite{Marta08-PRC} and are discussed in details in sec.~\ref{subsec:Relative}. Subsequently, also absolute cross section data for the four most important $\gamma$-transitions are derived from the spectra. This analysis is performed both by classical peak integrals for the addback mode data (sec.~\ref{subsec:Absolute}), and by $\gamma$-line shape analysis for the singles mode data (sec.~\ref{subsec:Lineshape}). 
Finally, by moving the detector to 19.5\,cm distance from the target, more precise branching ratios for the decay of the 259\,keV resonance are presented (sec.~\ref{subsec:Branchings}).

\subsection{Ratio of the cross sections for capture to the ground state and the 6792\,keV excited state in $^{15}$O}
\label{subsec:Relative}

% ================
\begin{figure*}[t!!]
\centering
% \scalebox{1.0}[0.35]{%
 \includegraphics[angle=0,width=1.0\textwidth]{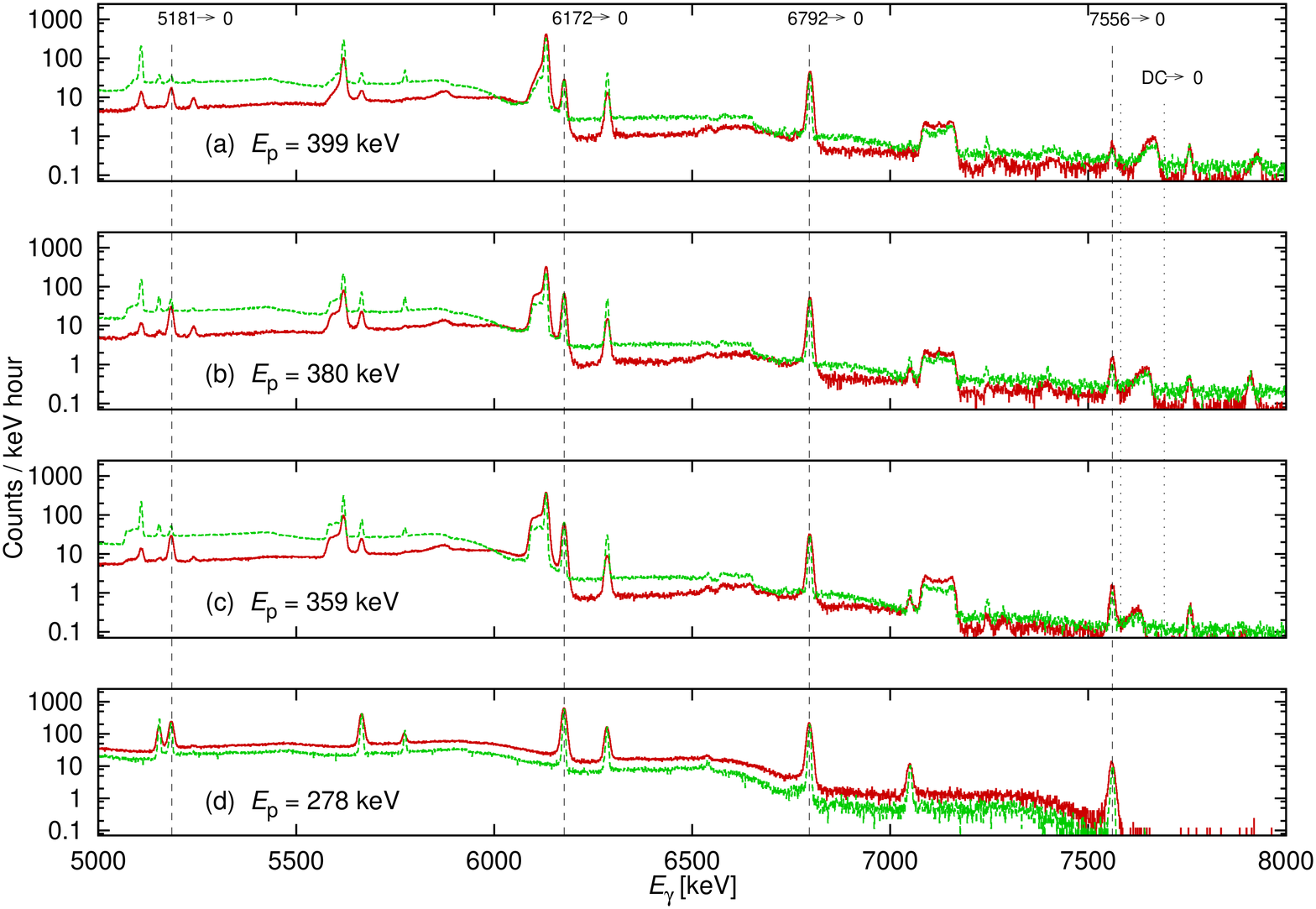}
 %}%scalebox
 \caption{\label{fig:Spectra_highenergy}(Color online) High-energy part of the $\gamma$-ray spectra for addback mode (red full line) and singles mode (green dashed line). (a)-(c): Off-resonant spectra at 
$E_{\rm p}$ = 399, 380, and 359\,keV, respectively, with the detector at 9.5\,cm distance from the target. (d): On-resonance spectrum 
on the $E_{\rm p}$ = 278\,keV resonance,
with the detector at 19.5\,cm distance from the target. The resonant contribution by the tail of the target is well visible also in the off-resonant spectra. The contaminant peaks stem from the $^{19}$F(p,$\alpha$$\gamma$)$^{16}$O reaction.}
\end{figure*}
% ================

For the relative analysis, the number of counts in the ground state capture peak at $E_\gamma$ $\approx$ 7600\,keV is compared with the number of counts in the secondary $\gamma$-ray at 6792\,keV (fig.~\ref{fig:Spectra_highenergy}). In such an analysis, only the relative uncertainty when extending the efficiency curve over this limited energy range contributes to the uncertainty of the ratio (0.8\% effect). 

\begin{figure*}[t]
\centering
%\resizebox{\textwidth}{!}{
%\scalebox{1.0}[0.7]{%
\includegraphics[angle=-90,width=\textwidth]{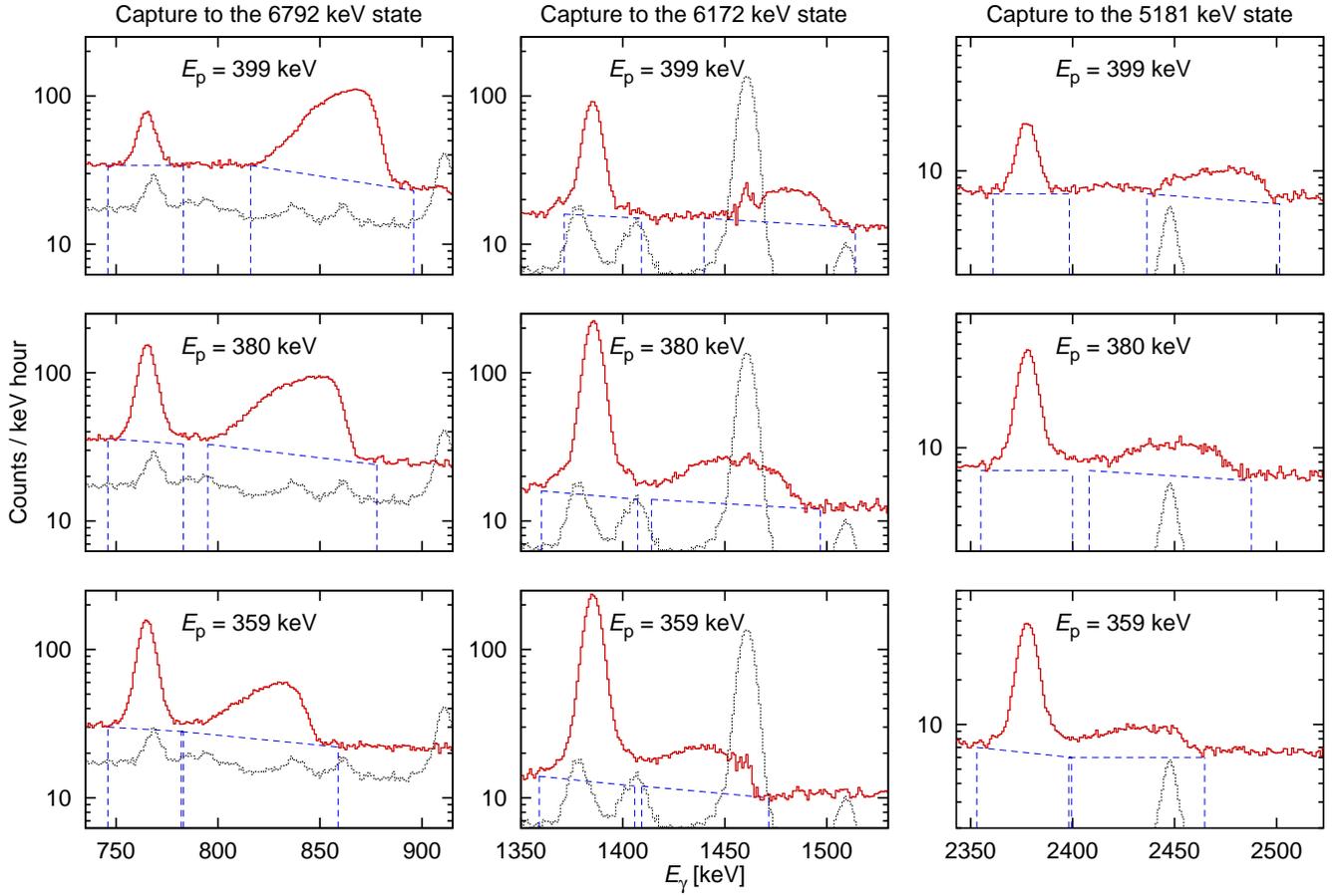}
%}
%}%resizebox
 \caption{\label{fig:Spectra_lowenergy}(Color online) 
Low-energy part of the $\gamma$-ray spectra for addback mode (red full line), after subtraction of the laboratory background. For completeness, the laboratory background is also included (black dotted line). Rows from top to bottom:  $E_{\rm p}$ = 399, 380, 359\,keV. Columns from left to right: Primary $\gamma$-ray for capture to the excited state at 6792, 6172, 5181\,keV. The peak from resonant capture (by the tail of the target) is clearly visible at the left of each panel. The non-resonant capture has a shape reflecting the profile of the target, convoluted with the energy-dependent cross section. The analysis of the 6172 primary (central column) is hampered by the strong $^{40}$K laboratory background line; this is reflected in higher uncertainty for this transition. The regions of interest and the assumed background are shown by blue dashed lines.}
\end{figure*}

The 6792\,keV counting rate contains some on-resonant contribution. This is due to the 60\,keV (full width at half maximum) thick target. When the beam slows down to the strong resonance at $E_{\rm p}$ = 278\,keV, it still finds some TiN in the tail of the target. In order to correct for this effect, the primary $\gamma$-rays for capture to this level are analyzed, as well, and the 6792\,keV counting rate is rescaled with the resonant/off-resonant ratio as obtained from the low-energy primaries (fig.~\ref{fig:Spectra_lowenergy}). The reduction in 6792\,keV counting rate by the escape-suppression shield contributes 1.2\% to the final uncertainty, and the summing-out correction for this peak contributes 0.6\%. 

Based on these data, the ratio
\begin{equation}
\label{eq:Ratiodefinition}
R_{\rm GS/6792}(E) = \frac{\sigma_{\rm GS}(E)}{\sigma_{\rm 6792}(E)}
\end{equation}
has been calculated (table~\ref{tab:CrossSectionRatios}). The present data supersede the data published previously in abbreviated form \cite{Marta08-PRC}, due to an upgraded background determination (fig.~\ref{fig:Spectra_lowenergy}, blue dashed lines), described in section \ref{subsec:Absolute}. 
The ratio depends only on the counting rates for the $E_\gamma$ $\approx$ 7600\,keV ground state capture $\gamma$-ray, for the $E_\gamma$ = 6792\,keV $\gamma$-ray (corrected for resonant capture as described above), and on the ratio of the $\gamma$-detection efficiencies at $E_\gamma$ $\approx$ 7600 and 6792\,keV. For the ground state capture $\gamma$-ray, a summing-in correction of up to 30\% (4.3\%) for addback (singles) mode was taken into account (table~\ref{tab:CrossSectionRatios}, last column).

%%%%%%%%%%%%%%%%%%%%%%%%%%%%%%%%
 \begin{table}[tb]
\caption{\label{tab:CrossSectionRatios} Cross section ratio $R_{\rm GS/6792}(E)$ and relative uncertainty. The size of the summing-in correction is also given. The present data supersede Ref.~\cite{Marta08-PRC}, due to an improved background determination.}
\begin{ruledtabular}
\begin{tabular}{clcccc}
        $E$ [keV] & mode & $R_{\rm GS/6792}(E)$ &
        stat. & syst. & Summing-in \\
& & [10$^{-2}$] & \multicolumn{2}{c}{uncertainty}& correction \\
        \hline
315.3$\pm$1.3 & addback & 5.24 & 11\% & 5.4\% & 30\% \\
& singles & 5.22 & 15\% & 2.7\% & 4.3\% \\
333.1$\pm$1.0 & addback & 5.33 & 4.8\% & 3.9\% & 21\% \\
& singles & 5.58 & 11\% & 2.5\% & 3.4\% \\
353.3$\pm$1.0 & addback & 5.20 & 3.5\% & 3.5\% & 19\% \\
& singles & 5.43 & 8.0\% & 2.3\% & 3.2\%
\end{tabular}
\end{ruledtabular}
\end{table}
%%%%%%%%%%%%%%%%%%%%%%%%%%%%%%%%

When computing $R_{\rm GS/6792}(E)$, the current measurement and the target stoichiometry and profile cancel out, eliminating the major sources of uncertainty. Therefore, the relative analysis method allows to derive data with much better precision than for absolute data. The present relative data can then be rescaled with averaged data for the well-studied cross section for capture to the 6792\,keV state, and uniquely precise data for capture to the ground state can be obtained.

The effective interaction energies 
%db begin
have been determined for each $\gamma$-line with two methods: First, the centroid of the off-resonant primary $\gamma$-line has been used, taking into account the reaction $Q$-value and $\gamma$-level energies. Second, the average energy, weighted with the predicted counts from the known target profile and the expected energy dependence of the cross section from the R-matrix S-factor curve \cite{Imbriani05-EPJA}. The two values were never more than 2.6\,keV apart, and their average was adopted for each line. The results 
%db end
are slightly different for ground state capture and capture to the 6792\,keV state, because the S-factor curve from previous R-matrix fits has a different slope for these two transitions. Therefore, the average of the two values is adopted as effective energy to be connected with 
%db begin
the cross section ratio 
%db end
$R_{\rm GS/6792}(E)$, with the assigned 1$\sigma$ error bar covering both effective energy values.

For the relative data, the total systematic uncertainty is 3.5 - 5.4\% in addback mode (table~\ref{tab:Uncertainties}). For singles mode, due to the lower summing corrections, it is 2.3 - 2.7\%.

For all three data points, the addback and singles mode data are in good agreement. Due to the higher $\gamma$-efficiency of the addback mode data (which, in turn, is due to the well-known addback factor of Clover-type detectors \cite{Duchene99-NIMA}, which has been redetermined for the present detector and geometry \cite{Szucs10-EPJA}) and due to the background reduction achieved by the escape-suppression shield for the addback mode data, the addback data have much better statistics than the singles mode. Therefore, the addback data are adopted for the further analysis despite their slightly higher systematic uncertainty.

%%%%%%%%%%%%%%%%%%%%%%%%%%%%%%%%
%%%%%%%%%%%%%%%%%%%%%%%%%%%%%%%%

%%%%%%%%%%%%%%%%%%%%%%%%%%%%%%%%
\subsection{Absolute cross sections based on the peak integrals of the addback mode data}
\label{subsec:Absolute}

As a second step, the absolute cross section for capture to the excited states at 5181, 6172, and 6792\,keV and to the ground state of $^{15}$O has been derived, accepting that the systematic uncertainty (table~\ref{tab:Uncertainties}) includes now also the contributions from current measurement, target stoichiometry and profile, and absolute detection efficiency. Only the addback mode data were considered.

In order to obtain the net counting rate, a straight-line background based on two flat regions to the left and right of the region of interest (ROI) has been subtracted from the integral over the ROI. This procedure was applied for every secondary except for the decay of the 6172\,keV excited state, where a different method was applied. It was repeated for each transition of the run at 399\,keV, both for the primary (resonant and non-resonant) and secondary $\gamma$-rays.

However, in many cases it was not possible to apply this method of background determination: At $E_{\rm p}$ = 359 and 380\,keV, the off-resonant part of the primaries lie close to the resonant peak (fig.~\ref{fig:Spectra_lowenergy}, second and third row). 
The secondary at $E_\gamma$ $\approx$ 6172\,keV was problematic, as well, due to the $^{19}$F(p,$\alpha$$\gamma$)$^{16}$O background peak at $E_\gamma$ $\approx$ 6130\,keV (fig.~\ref{fig:Spectra_highenergy}). For these spectra, a different method was instead used to estimate the background: The ratio between the difference in average counts per channel observed to the left and right of the peak, and the net area of the peak itself, was calculated. The ratios observed on the resonance, where no additional resonant contribution exists and where beam-induced background is negligible, have then been used to calculate the background at the same $\gamma$-energy in the problematic spectra.
For those problematic spectra, a minimum uncertainty of 5\% has been assumed for the quantity subtracted from the raw integral of the ROI. 
Finally, it was ensured that the 1$\sigma$ uncertainty of the counts includes also results with different choices of background regions. 

The net counting rate was then determined from the secondary $\gamma$-ray, rescaled for its non-resonant/resonant contributions determined by the primary $\gamma$-rays. Based on the counting rate, the target stoichiometry and profile (sec.~\ref{subsec:Target}, fig.~\ref{fig:Clover_targetscans}), the beam current measurement, and the $\gamma$-detection efficiency (fig.~\ref{fig:Efficiency}), the cross section was calculated for these transitions. The angular distribution was assumed to exhibit negligible contributions from all Legendre polynomials except for zero and second order. The second order Legendre polynomial cancels out at the present detection angle of 55$^\circ$.

For the determination of the astrophysical S-factor from a single data point, it is necessary to make some assumption on the relative shape of the S-factor curve. For the present analysis, the S-factor was assumed to vary over the target thickness as given by the previous LUNA R-matrix curve \cite{Imbriani05-EPJA}. In order to check the uncertainty introduced by this assumption, the present analysis was repeated assuming a flat S-factor, and the full difference (1-9\%, depending on the transition and beam energy) was adopted as systematic uncertainty. The effective interaction energy \cite{Rolfs88-Book} was calculated based on the known target profile and the assumed S-factor behaviour. 
The uncertainties are half of the difference obtained by using a flat S-factor instead of the LUNA's curve \cite{Imbriani05-EPJA}.
%mm The effective interaction energies are slightly different for the different transitions because of the different slopes of the S-factor curves. 

%%%%%%%%%%%%%%%%%%%%%%%%%%%%%%%%
 \begin{table}[t!]
\caption{\label{tab:Uncertainties} Systematic uncertainties affecting cross section ratios ("relative", sec.~\ref{subsec:Relative}) and absolute cross sections ("absolute", secs.~\ref{subsec:Absolute} and \ref{subsec:Lineshape}) for addback mode.} 
\begin{ruledtabular}
%\resizebox{\columnwidth}{!}{
\begin{tabular}{lllr}
\multicolumn{2}{c}{Affecting data...} & Description & Amount \\ \hline
Relative & Absolute & Summing-in, ground state line & 3-5\% \\
Relative & Absolute & Escape-suppression efficiency & 1.2\% \\
Relative & Absolute & Slope of $\gamma$-efficiency curve & 0.8\% \\
Relative & Absolute & Summing-out & 0.6\% \\
& Absolute & Target, original stoichiometry & 6\% \\
& Absolute & Target, profile change & 5\% \\
& Absolute & Assumption on S-factor slope & 1-9\% \\
& Absolute & Beam current reproducibility & 2\% \\
& Absolute & Normalization of $\gamma$-efficiency & 1.8\% \\ \hline
Relative & & Total, addback mode & 3.5-5.4\% \\
& Absolute & Total, addback mode & 9-12\% \\
\end{tabular}
%}%resizebox
\end{ruledtabular}
\end{table}
%%%%%%%%%%%%%%%%%%%%%%%%%%%%%%%%

\subsection{Absolute cross sections based on the $\gamma$-line shape analysis of the singles mode data}
\label{subsec:Lineshape}

Subsequently, the absolute cross section for capture to the excited states at 5181, 6172, and 6792\,keV and to the ground state of $^{15}$O has also been calculated based on the $\gamma$-line shape analysis approach. To this end, only the singles mode data, which are essentially free from summing corrections, have been used. This approach is thus complementary to the one described in the previous section, which calculated peak integrals and used only the addback mode data. 

The $\gamma$-line shape analysis method has been described previously in details \cite{Formicola03-NIMA,Formicola04-PLB,Imbriani05-EPJA}, so it will only be outlined here. The analysis of the line shape of the primary $\gamma$-ray is possible because the observed
line shape of a primary transition is determined by the cross
section behavior $\sigma(E)$ in the proton energy interval
spanned by the incident beam during the slowing-down process in
the target. Each center-of-mass beam energy $E$ (at
which the reaction takes place) corresponds to a  $\gamma$-ray
energy 
\begin{equation} \label{eq:egamma}
E_\gamma = E+Q-E_{\rm x}+\Delta E_{\rm Doppler} - \Delta E_{\rm Recoil}, 
\end{equation}
with $Q$ the reaction $Q$-value, $E_{\rm x}$ the energy of the excited state, and $\Delta E_{\rm Doppler/Recoil}$ the appropriate Doppler and recoil corrections. The $\gamma$-line shape is also
influenced by the energy loss of the protons in the target,
because the stopping power of the protons in titanium nitride is a function of
proton energy \cite{SRIM08.02}. 

The number of counts N$_i$ in channel i of the $\gamma$-spectrum, corresponding to the energy bin \mbox{[$E_{\gamma i}$,$E_{\gamma i}$+$\delta E_\gamma$]}, where $\delta E$ is the dispersion in units of keV per channel, is given by the expression
\begin{equation}\label{eq:lineshape_cts}
N_{i}~=~ \frac{\sigma(E_{i})\delta E_{\gamma}\eta_{fe}(E_{\gamma
i})b_k } {\varepsilon(E_{i})}
\end{equation}
for $E_{i}\leq E$. Here $E_{i}$ is the center-of-mass proton energy corresponding to channel $i$, $E$ is the incident proton energy in the center-of-mass, $\sigma(E_{i})$ is the cross section under study, $\eta_{fe}(E_{\gamma,i})$ is the $\gamma$-ray detection efficiency, $\varepsilon(E_i)$ is the
stopping power and b$_k$ is the branching of the transition under study.
The conversion from E$_{\gamma,i}$ to E$_{i}$ includes the Doppler and recoil effects, as shown in eq.\,(\ref{eq:egamma}). The resulting count rate is folded with the known energy resolution $\Delta E_\gamma$ of the $\gamma$-ray detector to obtain the experimental
line-shape.

%%%%%%%%%%%%%%%%%%%%
\begin{figure*} [tb]
\begin{center}
\includegraphics[width=\textwidth]{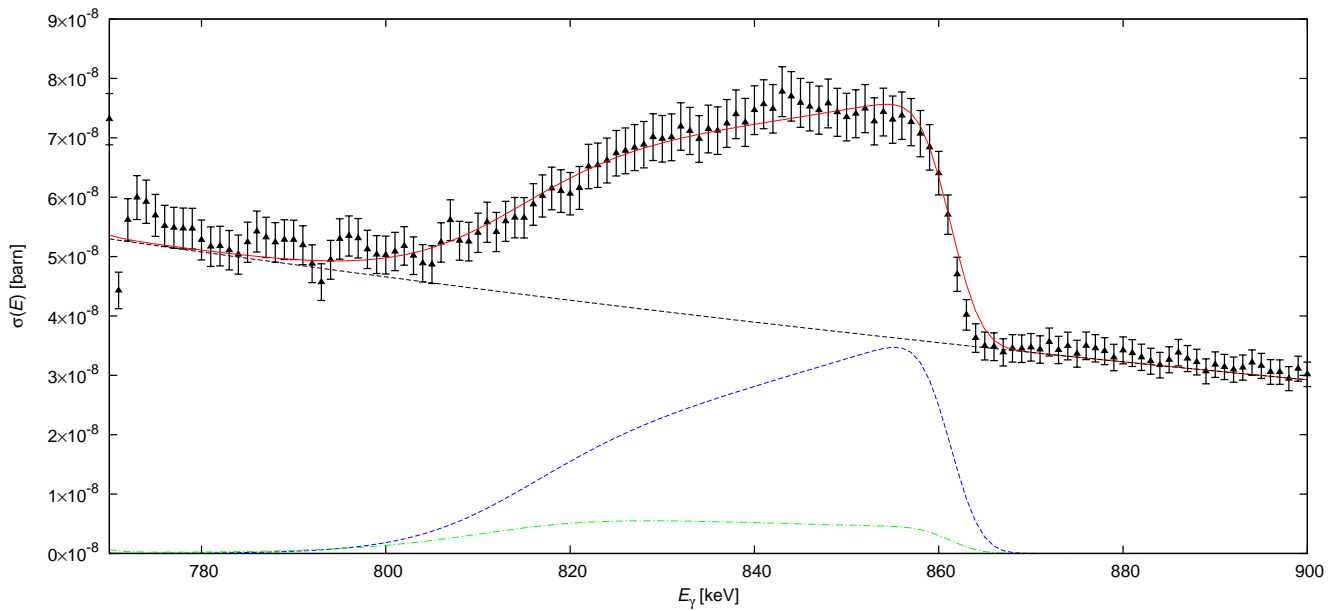}%}
\caption{(Color online) Typical $\gamma$-ray line shape obtained at $E_{\rm p}$ = 380\,keV, in singles mode, for capture to the 6792\,keV state. The dashed black line is the assumed background, fitted outside the peak area. The dash-dotted (green)
line corresponds to the expected resonant contribution (first part of eq.\,(\ref{eq:sigmatot})), and the dotted (blue) line to the fitted non resonant part (second part of eq.\,(\ref{eq:sigmatot})). The solid (red) line is the sum of these two components and the background.} \label{fig:lineshape_ep380_679}
\end{center}
\end{figure*}
%%%%%%%%%%%%%%%%%%%%

To facilitate the fit, the cross section $\sigma(E)$ entering into eq.\,(\ref{eq:lineshape_cts}) is then parameterized, in the limited energy window defined by the target thickness $\Delta E_{\rm Target}$, as the sum of a resonant term described by the Breit-Wigner formula, and a non-resonant term, for which a constant astrophysical S-factor $S_{\rm nr}$ is assumed:
\begin{equation}\label{eq:sigmatot}
\sigma(E_{i})=
\frac{\lambda^2}{\pi}\omega\gamma\frac{\Gamma}{(E_{i}-E_{\rm R})^2+(\Gamma/2)^2)}
+ \frac{S_{\rm nr}e^{-2\pi\eta}}{E_{i}}
\end{equation}
Here, $\lambda$ is the de Broglie wavelength, $\omega\gamma$ the strength value of the 259\,keV resonance (here, 12.9\,meV was used \cite{Imbriani05-EPJA}, very close to the recently recommended value of 13.1\,meV \cite{Seattle09-workshop}), $E_{\rm R}$ the energy of the resonance, $\Gamma$ the energy-dependent total width of the resonance, and $e^{-2\pi\eta}$ is the Sommerfeld parameter. Since the branching ratios and the $\omega\gamma$ of the resonance are kept fixed, the free parameters in this procedure are the non-resonant S-factor $S_{\rm nr}$, the background parameters and the energy of the beam. They are fitted to best reflect the shape of the primary $\gamma$-line by reducing the $\chi^{2}$.
After the fit has converged, the cross section $\sigma$ under study here is given by the average of the $\sigma(E_{i})$ values, weighted for their contribution to the total statistics. 

Figure \ref{fig:lineshape_ep380_679} shows a typical case for the primary $\gamma$-ray spectrum, together with the fit described above. The drop in the $\gamma$-ray yield towards lower
energies reflects mainly the drop of the cross section due to the lower Coulomb barrier penetrability at lower energy. The energy of the high energy edge of the peak provides an independent cross-check on the assumed beam energy from the accelerator energy calibration \cite{Formicola03-NIMA}.
Possible variations of the stoichiometry of the titanium nitride target during the beam bombardment have been monitored as described above (sec.\,\ref{subsec:Target} and fig.\,\ref{fig:Clover_targetscans}). 

The final astrophysical S-factor obtained from the line-shape analysis described in the present section was found to be in excellent agreement with the data from the peak-integral approach described in the previous section. It should be noted that while the present line-shape analysis is based on the singles mode spectra, the peak integral analysis is based on the addback mode data. The agreement between these two approaches confirms their reliability.

\begin{figure*}[t]
\centering
\includegraphics[angle=0,width=0.5\textwidth]{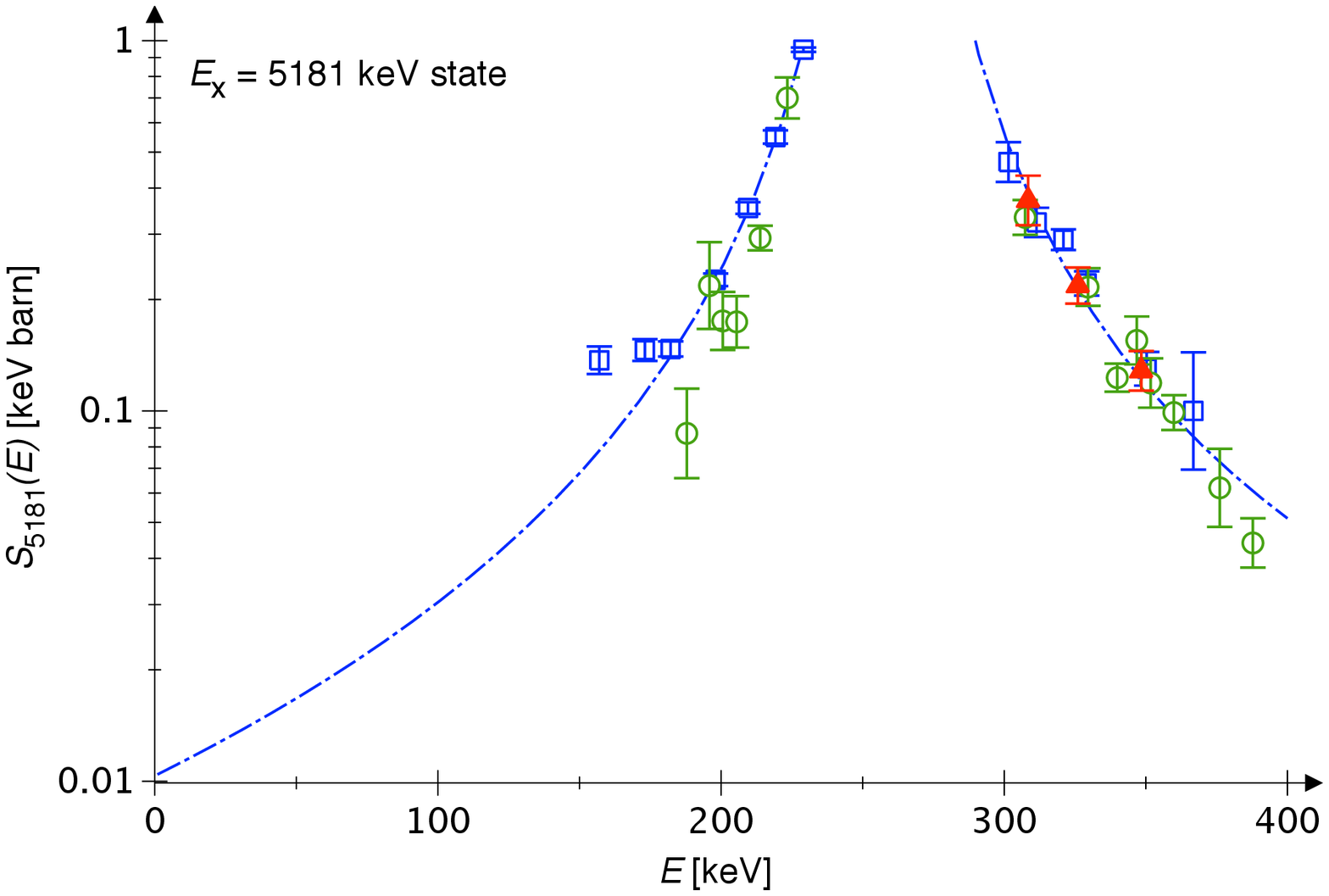}%
\includegraphics[angle=0,width=0.5\textwidth]{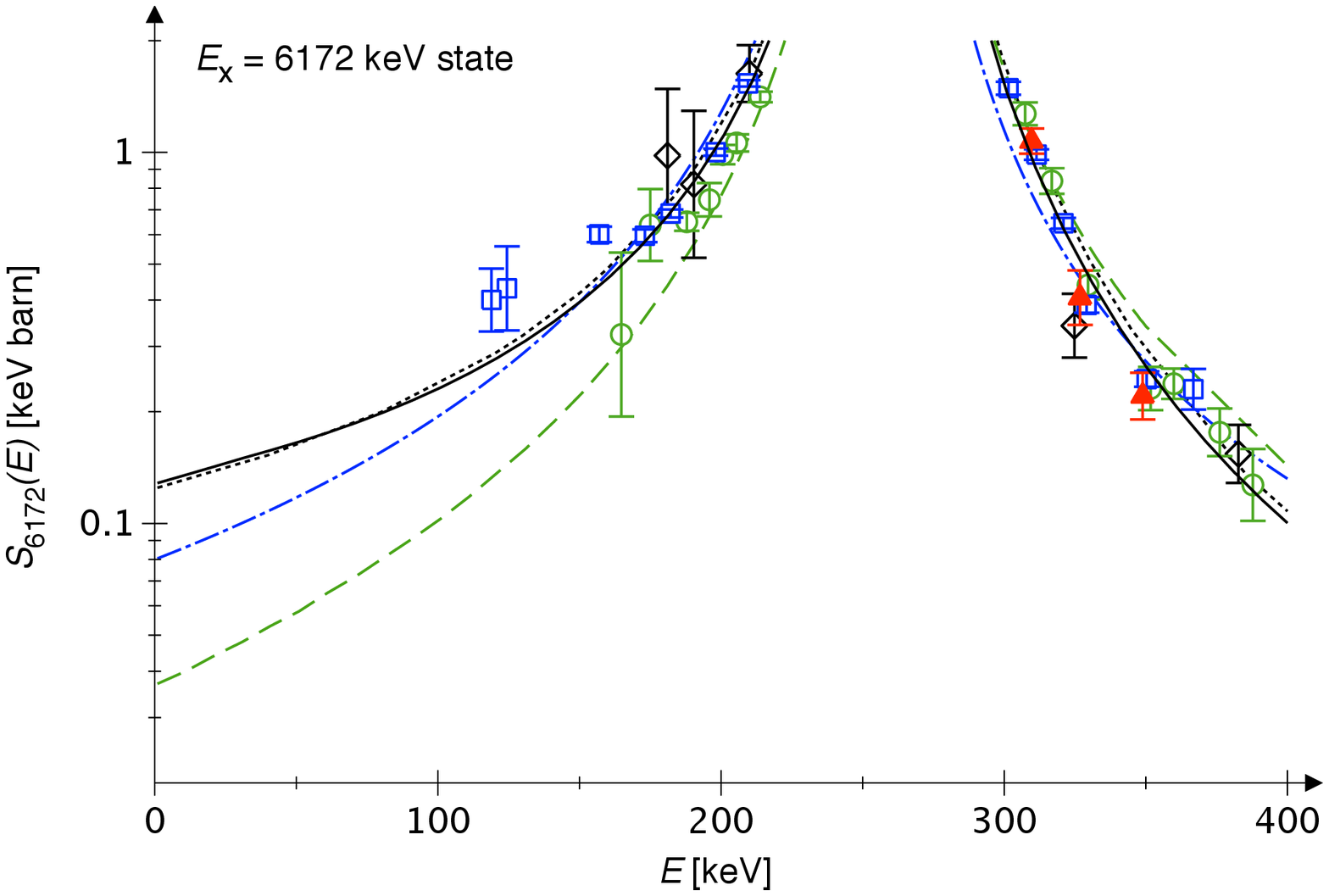}\\
\includegraphics[angle=0,width=0.5\textwidth]{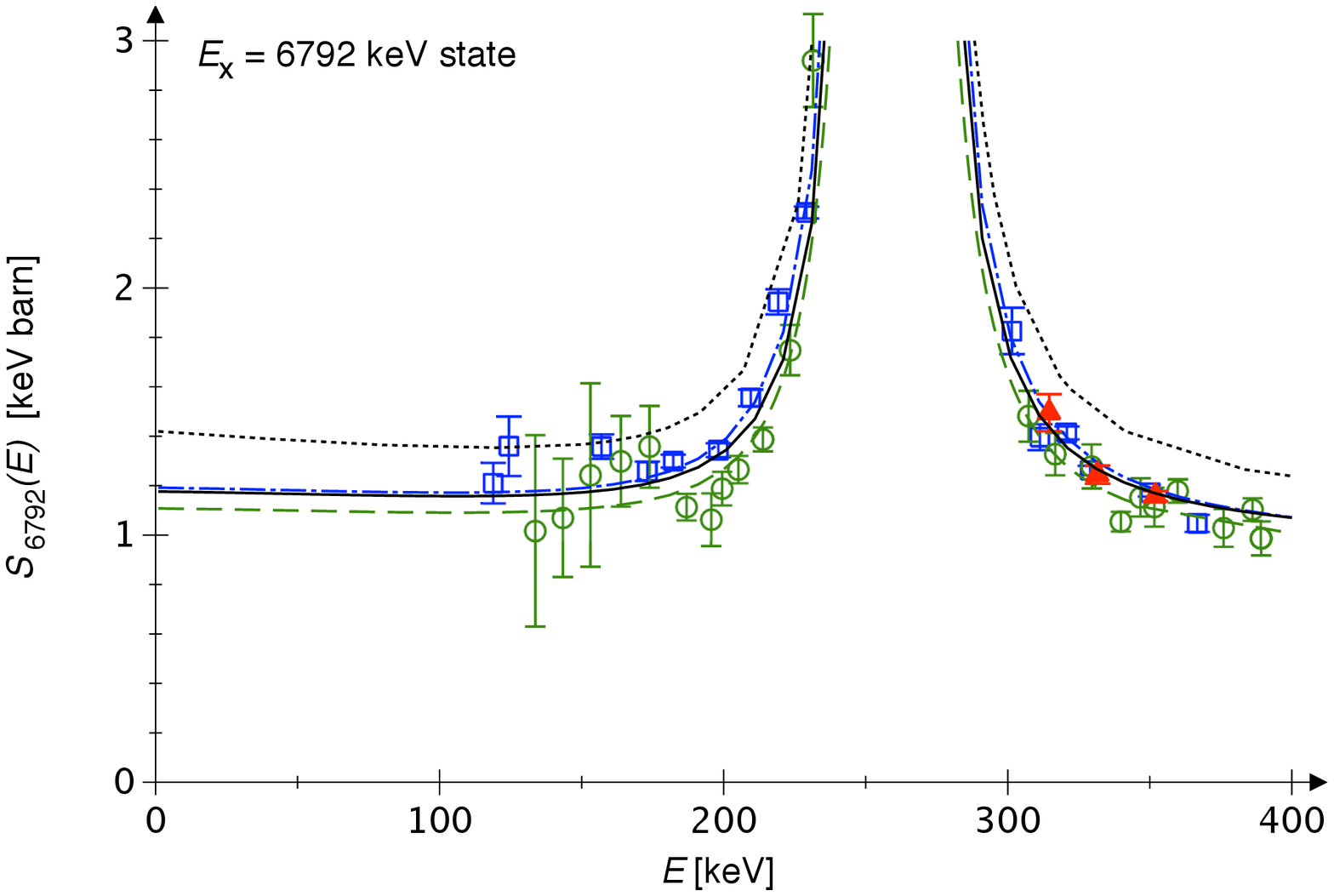}%
\includegraphics[angle=0,width=0.5\textwidth]{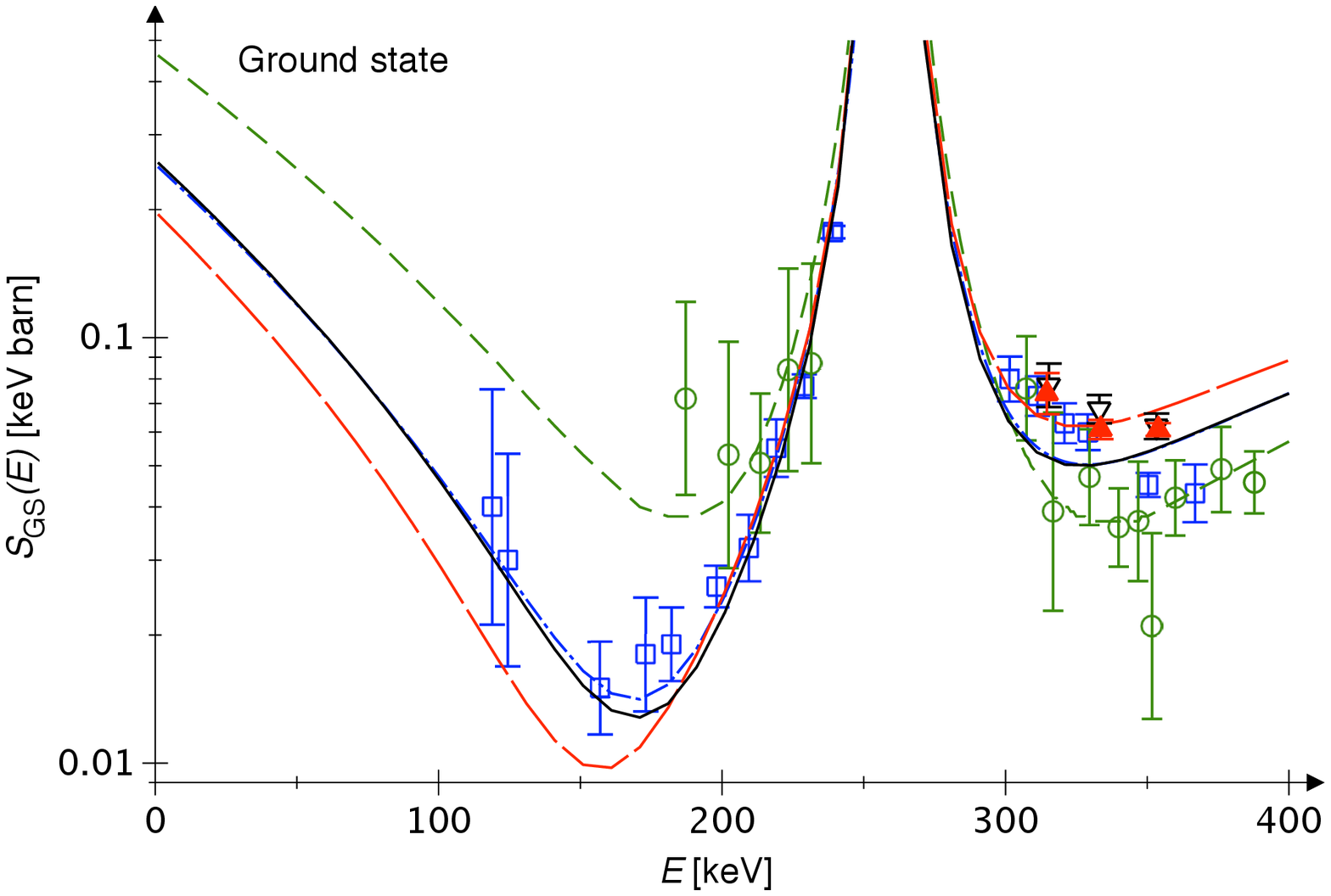}
 \caption{\label{fig:Sfactor}(Color online) 
S-factor for capture to the excited states at 5181, 6172, and 6792\,keV and to the ground state in $^{15}$O. Data: Black diamonds \cite{Schroeder87-NPA}, blue squares \cite{Formicola04-PLB,Imbriani05-EPJA}, green circles \cite{Runkle05-PRL}, red full triangles (present work, average of secs.~\ref{subsec:Absolute} and \ref{subsec:Lineshape}). R-matrix fits: Black dotted curve \cite{Mukhamedzhanov03-PRC}, blue dash-dotted curve \cite{Formicola04-PLB,Imbriani05-EPJA}, green dashed curve \cite{Runkle05-PRL}, black full curve \cite{Seattle09-workshop}, red long-dashed curve \cite{Marta08-PRC}. --- For ground state capture, the black inverted triangles represent the present relative data (sec.~\ref{subsec:Relative}), rescaled with the averaged S-factor for capture to the 6792\,keV state as described in the text. For capture to the 5181\,keV state, no R-matrix fits are given in Refs.~\cite{Mukhamedzhanov03-PRC,Runkle05-PRL}. Error bars reflect the statistical uncertainty.
 }
\end{figure*}

The final S-factor values from the present experiment are obtained by forming the simple average value of the two approaches (secs.\,\ref{subsec:Absolute} and \ref{subsec:Lineshape}). The data are summarized in table\,\ref{tab:Sfactors} and plotted in fig.\,\ref{fig:Sfactor}. 
%gi The systematic uncertainties are given in table\,\ref{tab:Uncertainties}.

%%%%%%%%%%%%%%%%%%%%%%%%%%%%%%%%
% ========================
\begin{table*}[bt]
\caption{\label{tab:Sfactors} S-factor results for capture to the ground state and to the excited states at 5181, 6172, and 6792\,keV. The effective energy $E$ is given in keV, the S-factor $S$ in keV barn, and the relative uncertainties for $S$ in percent.}
\begin{ruledtabular}
\resizebox{\textwidth}{!}{
\begin{tabular}{|*{4}{crrr|}}
\multicolumn{4}{|c|}{Capture to ground state} & \multicolumn{4}{c|}{Capture to 5181\,keV state} & \multicolumn{4}{c|}{Capture to 6172\,keV state} & \multicolumn{4}{c|}{Capture to 6792\,keV state} \\
$E$	&	$S_{\rm GS}$	&	$\Delta_{\rm stat}$	&	$\Delta_{\rm syst}$	&	$E$	&	$S_{5181}$	&	$\Delta_{\rm stat}$	&	$\Delta_{\rm syst}$	&	$E$	&	$S_{6172}$	&	$\Delta_{\rm stat}$	&	$\Delta_{\rm syst}$	&	$E$	&	$S_{6792}$	&	$\Delta_{\rm stat}$	&	$\Delta_{\rm syst}$	\\	\hline
314.6$\pm$1.0	&	0.074	&	11\%	&	12\%	&	310.6$\pm$2.2	&	0.370	&	16\%	&	11\%	&	310.5$\pm$1.0	&	1.072	&	8\%	&	12\%	&	315.9$\pm$1.3 	&	1.495	&	5.0\%	&	9\%	\\	
333.6$\pm$1.0	&	0.061	&	5\%	&	11\%	&	327.6$\pm$1.6	&	0.218	&	12\%	&	12\%	&	326.6$\pm$1.0	&	0.406	&	18\%	&	12\%	&	332.6$\pm$1.0 	&	1.245	&	3.0\%	&	9\%	\\	
353.9$\pm$1.0	&	0.061	&	4\%	&	10\%	&	350.9$\pm$2.5	&	0.128	&	13\%	&	10\%	&	351.1$\pm$2.2	&	0.220	&	15\%	&	10\%	&	352.7$\pm$1.0	&	1.157	&	1.7\%	&	9\%	\\	
\end{tabular}
}%resizebox
\end{ruledtabular}
\end{table*}

%%%%%%%%%%%%%%%%%%%%%%%%%
\subsection{Branching ratios for the decay of the 259\,keV resonance, obtained in far distance}
\label{subsec:Branchings}

In order to determine the branching ratios for the decay of the 259\,keV $\frac{1}{2}^+$ resonance ($E_{\rm x}$ = 7556\,keV in $^{15}$O), the Clover detector was moved to a farther geometry, with its front face at 19.5\,cm distance from the target position, again at an angle of 55$^\circ$ with respect to the beam direction. For the branching ratio analysis, both addback and singles mode data have been analyzed and were found to agree within their statistical uncertainty in all cases. In the following text, only the singles mode data will be discussed. 

The detection efficiency was again established as described above (sec.~\ref{subsec:GammaDetection}), with an analogous quality of the efficiency curve as the one shown for the 9.5\,cm geometry (fig.~\ref{fig:Efficiency}). It should be noted that the efficiency curve does not depend on the branching ratios, just on the assumption of 1:1 cascade ratios without feeding or intermediate decay corrections for the two transitions through the states at 6172 and 6792\,keV, and on the assumption of isotropy \cite{Povh59-PR}. 

For the determination of the decay branchings of the 259\,keV resonance, only the secondary $\gamma$-rays at 5181, 5241, 6172, and 6792\,keV and the ground state primary $\gamma$-ray at 7556\,keV were used (fig.~\ref{fig:Spectra_highenergy}, bottom panel). Therefore only the relative $\gamma$-efficiency in the limited energy range 5181-7556\,keV is needed. Owing to the good quality of the $\gamma$-efficiency curve, over this limited energy range the efficiencies relative to the 6172\,keV normalization point are known on the level of $\pm$0.5\%, enabling a precise determination of the branching ratios. 

For the major transitions through the excited states at 5181, 6172, and 6792\,keV, the present branching ratios (tab.~\ref{tab:Branchings}) are in excellent agreement with the modern literature \cite{Runkle05-PRL,Imbriani05-EPJA}. However, some minor discrepancies arise when it comes to the minor transitions. 

The ground state transition has been the subject of discussion in recent years. It is now well-known that the previously accepted value of (3.5$\pm$0.5)\% \cite{Tabata60-JPSJ,Hebbard63-NP,Ajzenberg13-15} was much too high, probably due to summing-in. The two most recent previous branching ratio measurements \cite{Runkle05-PRL,Imbriani05-EPJA} were both performed at about 20\,cm distance, where there is still more than 10\% summing-in correction.
The present value of (1.49$\pm$0.04)\% has been obtained at 19.5\,cm distance, with just 2.0\% summing-in correction for the singles mode data, much less than in previous works. 
Note that the value (1.53$\pm$0.06)\% from an abbreviated version of the present work \cite{Marta08-PRC} had been obtained in closer geometry, at 9.5\,cm distance, with 7.4\% summing-in correction. 
The present (1.49$\pm$0.04)\% ground state branching supersedes all previous LUNA branching ratio measurements of the 259\,keV resonance, i.e. \cite{Imbriani05-EPJA,Marta08-PRC}.

For the transition to the 5241\,keV state, the previous (0.6$\pm$0.3)\% value \cite{Imbriani05-EPJA} was possibly affected by feeding through higher-lying excited states. Based on the difference between 5241 $\rightarrow$ 0 and 7556 $\rightarrow$ 5241 $\gamma$-rays, this feeding contribution amounts to (0.20$\pm$0.10)\% of the total decay branching. It is probably due to the 6859\,keV state, which decays to 100\% to the 5241\,keV state \cite{Ajzenberg13-15}. However, such a weak feeding could possibly also arise through the 6172 or 6792\,keV states, so in absence of conclusive evidence this (0.20$\pm$0.10)\% is not assigned to any transition.

For the transition to the 5181\,keV state, the present data confirms the slightly higher modern values \cite{Runkle05-PRL,Imbriani05-EPJA} with respect to the compilation \cite{Ajzenberg13-15}.

% ==================
\begin{table}[tb]
\centering
\caption{Branching ratios for the decay of the 259\,keV resonance ($E_{\rm x}$ = 7556\,keV in $^{15}$O) obtained with the Clover detector in singles mode, at 19.5\,cm distance from the target. The numbers are compared with previous data \cite{Ajzenberg13-15,Runkle05-PRL,Imbriani05-EPJA}.}
\label{tab:Branchings}
\begin{tabular}{l@{$\rightarrow$}r*{4}{r@{$\pm$}l}}
%
%\hline 
\multicolumn{2}{l}{}  & \multicolumn{8}{c}{Branching [\%]}\\ \cline{3-10}
\multicolumn{2}{c}{} & \multicolumn{2}{l}{Ajzenberg-} & \multicolumn{2}{c}{TUNL \cite{Runkle05-PRL}} & \multicolumn{2}{l}{LUNA \cite{Imbriani05-EPJA}} & \multicolumn{2}{l}{LUNA,} \\
\multicolumn{2}{c}{} & \multicolumn{2}{l}{Selove \cite{Ajzenberg13-15}} & \multicolumn{2}{c}{} & \multicolumn{2}{l}{} & \multicolumn{2}{l}{present work} \\
\hline
7556 & 0 & 3.5&0.5 & 1.70&0.07 & 1.6&0.1 & 1.49&0.04 \\
& 5181 &  15.8&0.6 & 17.3&0.2 & 17.1&0.2 & 17.3&0.2 \\
& 5241 &  \multicolumn{4}{c}{} & 0.6&0.3 & 0.15&0.03 \\
& 6172 &  57.5&0.4 & 58.3&0.5 & 57.8&0.3 & 58.3&0.4 \\
& 6792 & 23.2&0.6 & 22.7&0.3 & 22.9&0.3 & 22.6&0.3 \\
\hline
\end{tabular}
\end{table}
% ==================

% ===========================================================
\section{R-matrix analysis of ground state capture}

For the purpose of an R-matrix analysis, the present relative data (sec.~\ref{subsec:Relative}, table~\ref{tab:CrossSectionRatios}) have been renormalized using a weighted average S-factor for capture to the 6792\,keV state. Based on these values and the corrected Schr\"oder data \cite{Schroeder87-NPA,Formicola04-PLB}, a new R-matrix fit for ground state capture has already been presented in the abbreviated form of the present work \cite{Marta08-PRC}. The present updated relative data are close to the values published in abbreviated form \cite{Marta08-PRC}, so this update does not warrant a revised fit.

Also the present absolute data (sec.~\ref{subsec:Absolute}) do not significantly deviate from the relative data, renormalized as stated above (fig.~\ref{fig:Sfactor}, bottom right panel). It should be noted that the present absolute data for capture to the 6792\,keV state (fig.~\ref{fig:Sfactor}, bottom left panel) are in excellent agreement with previous data \cite{Runkle05-PRL,Imbriani05-EPJA} and R-matrix fits \cite{Runkle05-PRL,Imbriani05-EPJA}, confirming that the renormalization procedure was adequate. By design the absolute data have higher uncertainty than the relative data (table~\ref{tab:Uncertainties}) that have already been included in the fit \cite{Marta08-PRC}, so no new R-matrix fit is attempted here. 

The previous fit \cite{Marta08-PRC} is instead shown again here (fig.~\ref{fig:Sfactor}, bottom right panel), leading to $S_{\rm GS}$(0) = 0.20$\pm$0.05\,keV\,barn. That value is lower than the recently recommended 0.27$\pm$0.05\,keV\,barn \cite{Seattle09-workshop}, but still in agreement given the error bars. The difference is mainly due to the fact that 
in the present work, only the present and the Schr\"oder \cite{Schroeder87-NPA} data (corrected for summing-in \cite{Formicola04-PLB}) are included. The data from Refs.\,\cite{Imbriani05-EPJA,Runkle05-PRL} are excluded due to concerns about the summing corrections. In Ref.\,\cite{Seattle09-workshop}, instead, the data from Refs.\,\cite{Imbriani05-EPJA,Runkle05-PRL} have also been included in the fit.
% db end

% ===========================================================
\section{Summary and outlook}

The $^{14}$N(p,$\gamma$)$^{15}$O reaction has been studied with a composite Clover-type detector at the LUNA underground facility at $E_{\rm p}$ = 359, 380, and 399\,keV, in an energy range important for future R-matrix fits of capture to the ground state in $^{15}$O. Precise cross section ratios for ground state capture relative to capture to the 6792\,keV state have been presented, updating and extending their previous abbreviated publication \cite{Marta08-PRC}. 

The present, precise relative cross section data (table~\ref{tab:CrossSectionRatios}) helped resolve the discrepancy between the previous, conflicting extrapolations for ground state capture \cite{Formicola04-PLB,Runkle05-PRL}, in favour of Ref.\,\cite{Formicola04-PLB}. The present recommended value of $S_{\rm GS}$(0) = 0.20$\pm$0.05\,keV\,barn is based on a dataset where the summing-in correction is not larger than 50\% \cite[]{Schroeder87-NPA} for the high-energy data and not larger than 30\% for the present, lower-energy data. 

The present absolute cross sections for capture to the excited states at 5181, 6172, and 6792\,keV (fig.~\ref{fig:Sfactor}, table~\ref{tab:Sfactors}) 
have been obtained with two independent analysing methods (secs.\,\ref{subsec:Absolute} and \ref{subsec:Lineshape}). They
are generally in good agreement with previous works \cite{Runkle05-PRL,Imbriani05-EPJA} and in some cases more precise. They are in overall good agreement with the most recent R-matrix fit \cite{Seattle09-workshop}. Because of their limited energy span, the present data alone cannot form the basis of new extrapolations. However, they may serve as useful reference points in an energy range that may be accessible not only at LUNA, but also at future underground accelerators. 

The new branching ratios for the decay of the 259\,keV resonance that are shown here improve the precision of the database for this resonance. Since this resonance is often used as normalization point for experimental work on the $^{14}$N(p,$\gamma$)$^{15}$O reaction \cite[e.g.]{Formicola04-PLB,Runkle05-PRL,Imbriani05-EPJA,Marta10-PRC}, this improved information again facilitates future precision studies of this reaction.

The present data are an important ingredient in updates of the standard solar model \cite{PenaGaray08-arxiv,Haxton08-ApJ,Serenelli09-ApJL}. When experimental data for the flux of solar CNO neutrinos due to the $\beta$-decay of $^{13}$N and $^{15}$O become available from Borexino \cite{Borexino08-PRL} or SNO+ \cite{Chen05-NPBPS}, precise $^{14}$N(p,$\gamma$)$^{15}$O cross sections may contribute to a direct measurement of the solar metallicity through a comparison of CNO and $^8$B neutrino fluxes \cite{Haxton08-ApJ}.

Possible next steps in improving the precision for the extrapolated S-factor of this reaction \cite{Seattle09-workshop} are to re-study the cross section at higher energies \cite{Marta10-PRC}, in order to improve the extrapolation, 
and a remeasurement of the strength of the 259\,keV resonance.

% ===========================================================
\begin{acknowledgments}
This work has been supported by INFN and in part by the EU (ILIAS-TA RII3-CT-2004-506222), OTKA (K68801 and NN 83261), and DFG (BE4100/2-1 and RO439/41-1). 
\end{acknowledgments}
% ===========================================================

\end{document}